\documentclass[desactivate]{aa} 

\usepackage[varg]{txfonts}
\usepackage{orcidlink}
\usepackage{graphicx}	% Including figure files
\usepackage{amsmath}	% Advanced maths commands
\usepackage{amssymb}	% Extra maths symbols
\usepackage{float}
\usepackage{cprotect}
\usepackage{amsmath}
\usepackage{graphicx}
\usepackage{mathtools}
\usepackage{hyperref}
\usepackage{multirow}
\usepackage{verbatim}
\usepackage{color, colortbl}
\usepackage{soul}
\usepackage{enumerate}
\usepackage{tabularx}
\usepackage{capt-of}

\usepackage{natbib}
\bibpunct{(}{)}{;}{a}{}{,}

\newcommand{\response}[1]{{#1}}% responses to referee
\newcommand{\responsetwo}[1]{{#1}}% responses to referee

\defcitealias{Polak2023}{Paper I}

\begin{document} 

   \title{Massive star cluster formation}

   \subtitle{ III. Early mass segregation during cluster assembly}
   % \subtitle{High Star Formation Efficiency with Resolution of Feedback from Individual Stars}

    \author{Brooke Polak
       \inst{1,2}\fnmsep\thanks{Fellow of the International Max Planck Research School for Astronomy and Cosmic Physics at the University of Heidelberg\\ \email{bpolak@amnh.org}}\orcidlink{0000-0001-5972-137X}
         \and
         Mordecai-Mark Mac Low\inst{2}\orcidlink{0000-0003-0064-4060}
         \and
        Ralf S. Klessen\inst{1,3,4,5}\orcidlink{0000-0002-0560-3172}
         \and
        Simon Portegies Zwart\inst{6}\orcidlink{0000-0001-5839-0302}
         \and
        Eric P. Andersson\inst{2}\orcidlink{0000-0003-3479-4606}
         \and
        Sabrina M. Appel \inst{2,7}\fnmsep\thanks{NSF Astronomy \& Astrophysics Postdoctoral Fellow}\orcidlink{0000-0002-6593-3800}
         \and
        Claude Cournoyer-Cloutier\inst{8}\orcidlink{0000-0002-6116-1014}
         \and
        Simon C. O. Glover\inst{1}\orcidlink{0000-0001-6708-1317}
        \and
        Stephen L. W. McMillan\inst{9}\orcidlink{0000-0001-9104-9675}
       }

\institute{Universit\"{a}t Heidelberg, Zentrum f\"{u}r Astronomie, Institut f\"{u}r Theoretische Astrophysik, Heidelberg, Germany
        \and
         Department of Astrophysics, American Museum of Natural History, New York, NY, USA
         \and
        Universit\"{a}t Heidelberg, Interdisziplin\"{a}res Zentrum f\"{u}r Wissenschaftliches Rechnen, Heidelberg, Germany
         \and 
         Harvard-Smithsonian Center for Astrophysics, Cambridge, MA, USA
         \and 
         Radcliffe Institute for Advanced Studies at Harvard University, Cambridge, MA, USA
         \and
        Sterrewacht Leiden, Leiden University, Leiden, the Netherlands
         \and
        Department of Physics and Astronomy, Rutgers University, Piscataway, NJ, USA
         \and
         Department of Physics and Astronomy, McMaster University, Hamilton, ON, Canada
         \and
         Department of Physics, Drexel University, Philadelphia, PA, USA
        }

   \date{Received 3 August 2024 / Accepted 18 February 2025}
 
  \abstract{

        Mass segregation is seen in many star clusters, but whether massive stars form in the center of a cluster or migrate there dynamically is still debated. 
        N-body simulations show that early dynamical mass segregation is possible when sub-clusters merge to form a dense core with a small crossing time. However, the effect of gas dynamics on both the formation and dynamics of the stars could inhibit the formation of the dense core. 
        We aim to study the dynamical mass segregation of star cluster models that include gas dynamics and self-consistently form stars from the dense substructure in the gas. Our models use the \textsc{torch} framework, which is based on \textsc{amuse} and includes stellar and magnetized gas dynamics, as well as stellar evolution and feedback from radiation, stellar winds, and supernovae. 
        Our models consist of three star clusters forming from initial turbulent spherical clouds of mass $10^4\,,10^5,\, 10^6\rm~M_\odot$ and radius $11.7\rm~pc$ that have final stellar masses of $3.6\times10^3\rm~M_\odot$, $6.5\times10^4\rm~M_\odot$, and $8.9\times10^5\rm~M_\odot$, respectively. There is no primordial mass segregation in the model by construction.
        All three clusters become dynamically mass segregated at early times via collapse confirming that this mechanism occurs within sub-clusters forming directly out of the dense substructure in the gas. The dynamics of the embedded gas and stellar feedback do not inhibit the collapse of the cluster.
        We find that each model cluster becomes mass segregated within 2~Myr of the onset of star formation, reaching the levels observed in young clusters in the Milky Way. However, we note that the exact values are highly time-variable during these early phases of evolution. 
        Massive stars that segregate to the center during core collapse are likely to be dynamically ejected, a process that can decrease the overall level of mass segregation again.

    }


   \maketitle

\section{Introduction}

Most stars form from giant molecular clouds (GMCs) in groups of tens to millions of stars called star clusters \citep[see, e.g.,][]{Lada2003ARA&A..41...57L, PortegiesZwart2010ARA&A..48..431P,Krause2020SSRv..216...64K}. Models suggest that star clusters form from the global hierarchical collapse of GMCs \citep{Semadeni2017MNRAS.467.1313V,Grudic2018MNRAS.481..688G}. As the GMC undergoes gravoturbulent collapse \citep{Larson1981MNRAS.194..809L}, fragmentation leads to the formation of dense star-forming clumps \citep{MacLowRevModPhys.76.125,Mckeedoi:10.1146/annurev.astro.45.051806.110602, Klessen2016SAAS...43...85K} called sub-clusters. These sub-clusters eventually merge if they are gravitationally bound, collapsing into a single star cluster. 

Many star clusters exhibit signs of mass segregation where the massive stars are concentrated in the center of the cluster \citep{Hillenbrand1998ApJ...492..540H}. This can happen dynamically when massive stars migrate to the center via two-body interactions \citep{Spitzer1969ApJ...158L.139S}, or it can occur primordially when massive stars are preferentially formed in the center where the gas density is highest \cite[e.g.,][]{Klessen2000, Mckee2003ApJ...585..850M}. Determining whether and when one of these modes of mass segregation dominates is essential for understanding the formation of stars and assembly of star clusters; whether star formation is spatially distributed randomly or according to mass presents two entirely different pictures of star formation. 

If a young star cluster is observed to be mass segregated earlier than expected given the cluster's crossing time, this would necessitate primordial mass segregation. However, N-body models presented in \citet{Allison2009ApJ...700L..99A} suggest a mechanism that allows for clusters to become dynamically segregated at much earlier times. They suggest that the collapse\footnote{Not to be confused with the core collapse that occurs at late dynamical times after the star cluster has already assembled.} of the star cluster -- when sub-clusters merge and create a short-lived dense core -- facilitates efficient and early dynamical mass segregation. However, it is unclear whether sub-clusters forming self-consistently from a collapsing GMC would still merge in a way that creates this dense core. Furthermore, including gas dynamics could also inhibit the collapse of the cluster's central region. 

We examine the mass segregation in three star cluster models, first presented in \citet[][hereafter \citetalias{Polak2023}]{Polak2023}. These simulations \response{resolve} individual stars \response{down to 4~M$_\odot$} forming from an initially turbulent spherical cloud of gas. By construction, the stars in our model are not primordially mass segregated. Thus we specifically study the timescale on which dynamical mass segregation occurs. We find that this process becomes important earlier than proposed by \citet{Allison2009ApJ...700L..99A}\response{, as transient periods of mass segregation occur even before the global collapse of the cluster.} We confirm that sub-clusters forming self-consistently from the cloud's substructure do form the dense cores needed for early dynamical segregation, and the embedding gas does not inhibit the collapse necessary for this mechanism. 

We briefly introduce our three star cluster models in Section~\ref{section:methods}, and in Section~\ref{section:results} we present the progression of mass segregation over each cluster's lifetime. In Section~\ref{sec:discussion} we examine the expected timescales for dynamical mass segregation and compare the degree of mass segregation in our clusters to observations in order to address whether primordial mass segregation is necessary to explain the mass segregation observed in young Galactic clusters. We conclude in Section~\ref{section:conclusions}. 

\section{Methods}
\label{section:methods}

We perform simulations of star clusters forming from turbulent spherical clouds of gas with \textsc{torch}\footnote{Version used for this work: \href{https://bitbucket.org/torch-sf/torch/commits/tag/massive-cluster-1.0}{https://bitbucket.org/torch-sf/torch/commits/tag/massive-cluster-1.0}} \citep{wall2019, 2020Wall}. The \textsc{torch} framework couples the magnetohydrodynamics code \textsc{flash} \citep{flash}, the N-body code \textsc{petar} \citep{petar}, and the stellar evolution code \textsc{SeBa} \citep{seba} via the Astrophysical MUltipurpose Software Environment \citep[AMUSE; ][]{PORTEGIESZWART2009369amuse1,Portegies2013CoPhC.184..456Pamuse2, amuse, amusebook}. Together, these three codes handle the magnetized gas dynamics, individual star formation via sink particles, stellar dynamics, stellar evolution, and stellar feedback in the form of radiation, winds, and supernovae (SNe). We provide a brief review of the numerical methods used by \textsc{torch} in this section; for more details, see \citetalias{Polak2023}, \citet{wall2019}, and \citet{2020Wall}.

\textsc{flash} is an adaptive mesh refinement grid code. We refined the grid on temperature and pressure gradients as well as the Jeans length $\lambda_{\rm J}=(\pi c_s^2 / [G\rho])^{1/2}$ of the gas. We used the HLLD Riemann solver \citep{Miyoshi2005JCoPh.208..315M} with third-order piecewise parabolic method reconstruction \citep{Colella1984JCoPh..54..174C}. A multigrid Poisson gravity solver \citep{Ricker2008ApJS..176..293R} calculates the gas self-gravity, and a leapfrog gravity-bridge \citep{Fujii2007PASJ...59.1095F, wall2019, portegieszwart2020CNSNS..8505240P-bridge} calculates the gravitational interaction between the star particles and gas. 

Sub-grid star formation in \textsc{torch} is modeled with sink particles \citep{Federrath_2010}. Briefly described, when the local gas density exceeds the Jeans density at the maximum level of refinement ($\rho_{\rm sink}$), a sink particle is formed to collect the dense\response{, star-forming} gas which would otherwise be unresolved. 
\response{Physically, sink particles represent star-forming cores of dense gas.}
Following the \citet{Truelove1997ApJ...489L.179T} criterion to avoid artificial fragmentation, we set $\lambda_J=5\Delta x=2r_{\rm sink}$. Upon formation, the mass of the gas denser than $\rho_{\rm sink}$ and within the sink's accretion radius $r_{\rm sink}$ is removed from the domain and added to the sink particle's mass. 
Sink particles have a list of stellar masses randomly sampled from the \citet{2002Sci...295...82Kroupa} initial mass function (IMF), and the sinks form the \response{next} star on the list as soon as they accrete the required amount of mass. \response{Primordial mass segregation does not occur globally in our models because every sink has a high enough star formation rate that they all fully sample the IMF.}

\response{
The accelerations of sink particles have contributions from other sinks, gas, and stars. The gravitational acceleration from other sinks is calculated by direct summation. Star particle gravity is combined with the gas by a cloud-in-cell mapping of stellar masses onto the grid. The gravitational acceleration from the gas and stars is then calculated using the potential returned by the multigrid Poisson solver.
The acceleration on the gas due to sinks is calculated by a direct summation using the distance from the sink to each gas cell center. The sink particle masses are also mapped onto the gas cells and used to calculate the acceleration of sinks on star particles.
Sink particles are allowed to merge in \textsc{torch}, but with $\lesssim20$ sinks in our models this rarely occurs. The time between accretion and star formation is less than a timestep for the majority of these simulations, so the effect of sink dynamics on star formation is negligible.} 

Stars are placed in a uniform \response{random} distribution within the sink's accretion radius, thereby suppressing any primordial mass segregation within the sink volume. The star's velocity is set by adding the sink's velocity to a velocity component with the magnitude sampled from a Gaussian distribution with \response{variance} $\sigma=c_s$ and with the direction sampled isotropically. The local sound speed $c_s$ is calculated by averaging the sound speed of cold gas ($\leq100\rm~K$) within $2r_{\rm sink}$. The sink does not form stars if there is no cold gas in the sink's vicinity. 

The modes of stellar feedback implemented in \textsc{torch} are radiation, winds, and SNe \citep{2020Wall}. The stellar evolution code \textsc{SeBa} evolves star particles from the zero-age main sequence until their stellar death, and provides the stellar feedback properties for the routines in \textsc{flash}. Two frequency groups of radiation, both ionizing and non-ionizing ultraviolet, are propagated via the Fervent ray-tracing module \citep{FERVENT10.1093/mnras/stv1906}.

\begin{table}
\label{table:params}   
\caption{Simulation parameters.}   
\centering   
\begin{tabular}{l|ccc}          
\hline\hline                        
 & M4 & M5 & M6 \\   
\hline                                   
    $M_\mathrm{cloud}$ [M$_\odot$] & $10^4$ & $10^5$ & $10^6$ \\
    $\rho_{c},\Bar{\rho}$ [M$_\odot$ pc$^{-3}$] & 2.8, 1.5 & 28, 15 & 280, 150 \\
    $\Sigma\ {\rm [M_\odot pc^{-2}]}$ & 23.25 & 232.5 & 2,325 \\
    $\lambda_{\rm J}$ [pc] & 10.0 & 3.2 & 1.0 \\
    $t_\mathrm{ff}$ [Myr] & 6.7 & 2.1 & 0.67 \\ 
    $B_{0,z}\rm~[\mu G]$ & 0.185 & 1.85 & 18.5 \\        
\hline\hline                        
Shared parameters & Value & Units \\   
\hline 
    $R_\mathrm{cloud}$ & 11.7 & pc\\
    $R_\mathrm{box}$ & 20.0 & pc\\ 
    $\alpha_\mathrm{v}=E_{\rm kin}/|E_{\rm pot}|$ & 0.15 & -\\
    $\Delta$x$_\mathrm{min}$ & 0.3125 & pc\\
    $\Delta$x$_\mathrm{max}$ & 1.25 & pc\\
    $r_\mathrm{sink}$ & 0.78125 & pc\\
    $\rho_\mathrm{sink}$ & $8\times 10^{-21}$ & g cm$^{-3}$\\ 
    $M_\mathrm{sink}$ & 246 & M$_\odot$\\ 
    $M_\mathrm{feedback}$ & 20 & M$_\odot$\\ 
    $M_\mathrm{n-body}$ & 4 & M$_\odot$\\ 
    $M_\mathrm{IMF}$ & 0.08--100 & M$_\odot$\\ 
\hline  
\end{tabular}
\tablefoot{Initial simulation parameters. Rows: initial cloud mass, cloud central and average volume density, initial average column density, Jeans length $\lambda_J=$ at initial temperature and mean density, free-fall time of mean density, peak initial vertical magnetic field, initial cloud radius, half-width of box, virial parameter, minimum cell width, maximum cell width, sink radius, sink threshold density, approximate initial sink mass, minimum feedback star mass, agglomeration mass of low-mass stars, mass sampling range of Kroupa IMF.}
\end{table}

We use the M4, M5, and M6 simulations introduced in \citetalias{Polak2023} which have initial cloud masses $10^4,\, 10^5$, and $10^6 \rm\, M_\odot$, respectively. The varying and shared simulation properties of the initial clouds are listed in Table \ref{table:params}. We aimed to keep the clouds as similar as possible to investigate the isolated effect of cloud mass on star cluster properties. 

The radius of all three clouds is $R_{\rm cloud}=11.7\rm\, pc$, so the density of the clouds increases with mass. We set the initial density profile of the cloud to be a Gaussian \citep{Bate1995MNRAS.277..362B,Goodwin2004A&A...414..633G} where $\rho_{\rm edge}/\rho_{\rm c}=1/3$. Our boundary conditions allow gas to flow into or out of the domain to prevent vacuums forming at the edge (see \citetalias{Polak2023} for details). The initial turbulence of the gas is set by a \cite{Kolmogorov1941DoSSR..30..301K} velocity spectrum using the same random seed for each simulation. With the same cloud radius, this results in the same morphology and spatial distribution of forming dense cores and stars facilitating comparisons between the three clouds. We set the initial magnetic field $\vec{B} = B_z \hat{z}$ to be uniform in the $z$ direction and decrease radially with the mid-plane density $\rho(x,y,z=0)$ in the $x$-$y$ plane: 
$B_z(x,y) = B_{0,z} \exp[ - (x^2+y^2) \ln(3) / R_\mathrm{cloud}^2]$
where the values of $B_{0,z}$ are listed in Table~\ref{table:params}.

We made a few modifications to standard \textsc{torch} in order to feasibly accommodate the large number of stars formed in the $10^5\rm~M_\odot$ and $10^6\rm~M_\odot$ clouds. \citetalias{Polak2023} provides an in-depth description of the modifications and an analysis of their effect. The relevant alteration for this paper is that we agglomerate low-mass stars below $M_{\rm agg}=4\rm\, M_\odot$ into single star particles with summed masses $>4\rm\, M_\odot$, which reduces the number of star particles on the grid by $\approx90\%$. \response{We agglomerate the star particles that are $<4\rm\, M_\odot$ in the order they are sampled until their combined mass is $\geq 4\rm\, M_\odot$. The mass of agglomerated star particles can lie between 4--8~M$_\odot$\footnote{\response{The maximum agglomerate mass of $8\rm~M_\odot$ comes from the rare case that the agglomerate mass is just under $4\rm\, M_\odot$ and the next star to be agglomerated is also just under $4\rm\, M_\odot$. For example, if the agglomerate mass is $3.9\rm\, M_\odot$, it will be combined with the next star to form that is $<4\rm\, M_\odot$. If this next star is also $3.9\rm\, M_\odot$, then the agglomerated particle will have a mass of $7.8\rm\, M_\odot$}}, but on average they are $\sim 4.5\rm\, M_\odot$.} Without agglomeration, the N-body code would have to compute the dynamics of $>10^6$ stars in the case of M6. 

\section{Results}
\label{section:results}

\begin{figure*}[t]
\centering
    \includegraphics[width=0.95\textwidth]{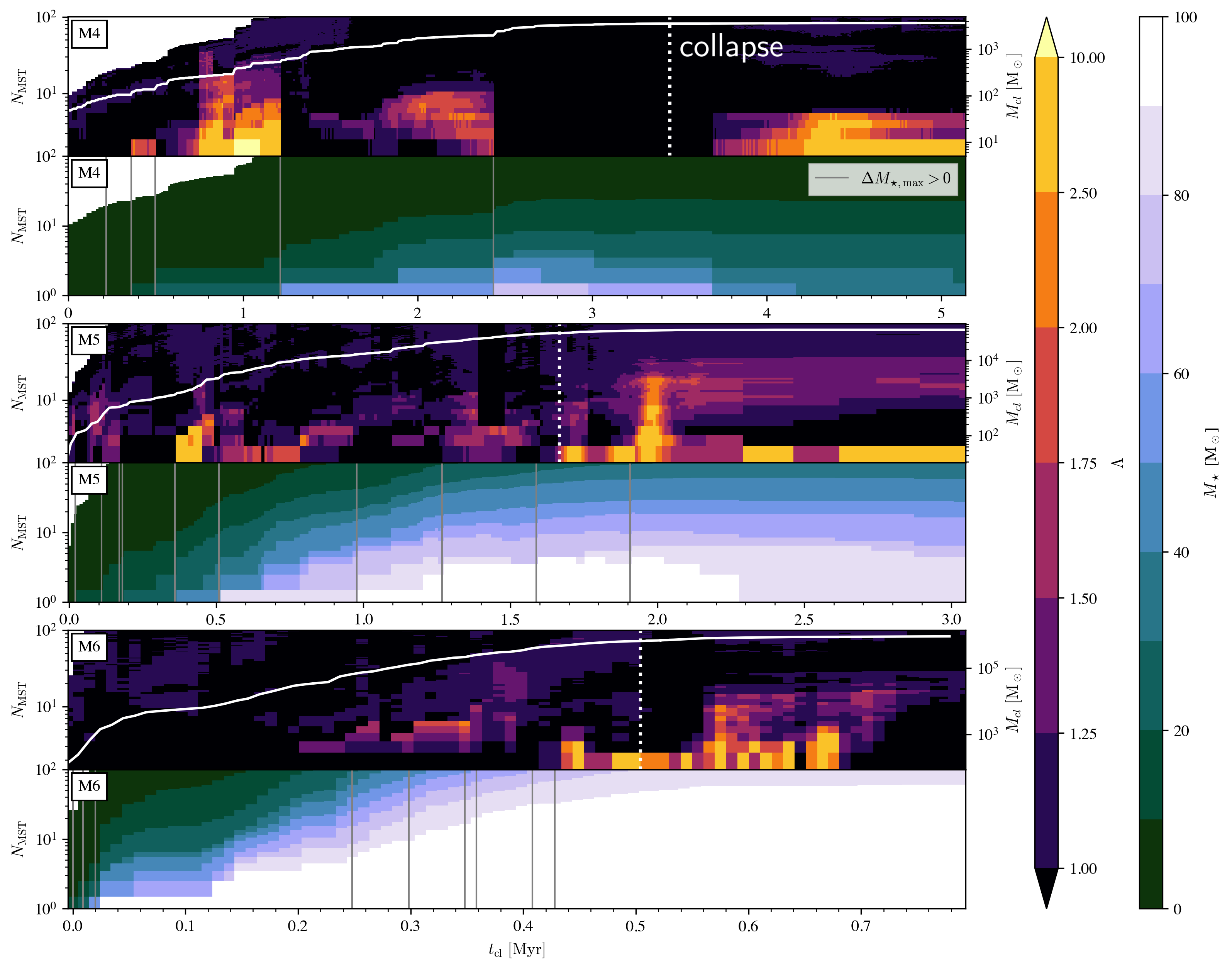}
    \caption{Mass segregation ratios $\Lambda$ and masses of $N_{\rm MST}$ most massive stars. The three rows with two grouped plots correspond to the M4, M5, and M6 models. The following descriptions apply to the two plots in each row. {\em Top}: Mass segregation ratio $\Lambda$ over time for the $N_{\rm MST}$ most massive stars in each cluster. The vertical dotted white lines indicate the time of collapse, where $R_{\rm rms}$ reaches a minimum. The solid white lines correspond to the right vertical axis showing the stellar mass of the cluster. {\em Bottom}: Mass of the $N^{\rm th}_{\rm MST}$ most massive star in each cluster. The grey vertical lines indicate the formation of a new most-massive star ($N_{\rm MST}=1$). As these are often not in the core, the time of formation, particularly in M4, corresponds to a drop in the apparent mass segregation. Note that each cluster was run to $\approx1.5t_\mathrm{ff}$, leading to different absolute timescales.}
    \label{fig:Nmst}
\end{figure*}

\begin{figure}[t]
\centering
    \includegraphics[width=1.0\hsize]{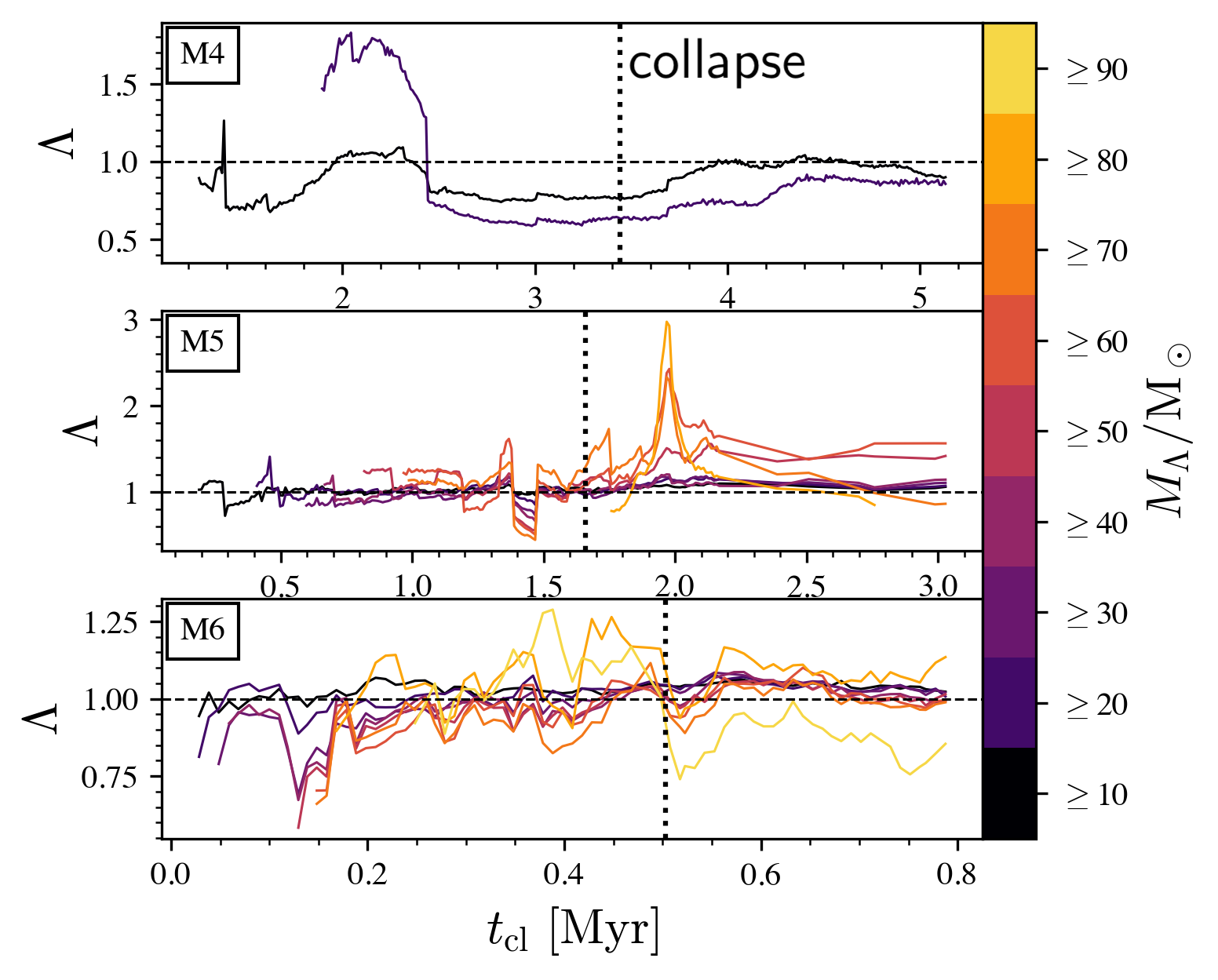}
    \caption{Mass segregation ratio $\Lambda$ over time for all stars above a given mass threshold $M_\Lambda$ shown on the color bar. $\Lambda$ is only calculated if there are $\geq5$ stars with mass $\geq M_\Lambda$. The threshold for mass segregation is $\Lambda > 1$, above the {\em horizontal dashed line}. The {\em vertical dotted lines} indicate the time of collapse, where $R_{\rm rms}$ reaches a minimum. Note the variable timescales in each panel.}
    \label{fig:msr}
\end{figure}

We quantified the degree of mass segregation in a star cluster using a minimum spanning tree (MST) method as described by \cite{10.1111/j.1365-2966.2009.14508.xAllison}. The basic idea is to use the total edge length of the MST formed by a group of stars to quantify their proximity to each other. This method compares the total edge length ($\ell_{\rm massive}$) of the MST made by the $N_\mathrm{MST}$ most massive stars in a cluster to the average total edge length ($\ell_{\rm rand}$) of $n_{\rm sample}$ MSTs formed by the same number of randomly selected stars. This gives a simple and comprehensive way of comparing the proximity of massive stars to that of all the stars in the cluster. This quantifies the mass segregation as a mass segregation ratio $\Lambda$, calculated by
\begin{equation}
    \mathrm{\Lambda=\frac{\langle \ell_{rand} \rangle}{\ell_{massive}} \pm \frac{\sigma_{rand}}{\ell_{massive}}}
\end{equation}
where $\sigma_{\rm rand}$ is the standard deviation of the $\ell_{\rm rand}$ values. Mass segregation is present when $\Lambda>1$. We use $n_{\rm sample}=500$ sets of randomly selected stars for calculating $\langle \ell_{\rm rand} \rangle \pm \sigma_{\rm rand}$. 

 \citet{Parker2015MNRAS.449.3381P} review the currently used measures of mass segregation and found the MST method to be the most accurate for measuring classical mass segregation, i.e., when massive stars are concentrated in particular regions of space. This review also finds that a random distribution of stars can result in values of $\Lambda=$1--2 for $N_{\rm MST}\le20$, and advise caution when interpreting values of $\Lambda$ below 2 for low $N_{\rm MST}$. However, they only used $N_\star=300$ stars in this test. The M5 and M6 clusters form $N_\star>15,000$ and $N_\star>150,000$ stars, respectively. The likelihood that the 20 most massive stars were all formed in the center of these clusters randomly is negligible. Therefore, for these clusters we consider all values of $\Lambda>1$ to signify mass segregation. The M4 cluster only forms $N_\star>500$ stars, so as a precaution we only consider mass segregation in M4 to be significant if $\Lambda>2$. We only consider bound stars for all the calculations in this study. 

In the top half of each row in Figure~\ref{fig:Nmst}, we plot the time evolution of the mass segregation ratio $\Lambda$ for the $N_{\rm MST} \leq 100$ most massive stars in each cluster. The bottom half of each row shows the mass of the $N_{\rm MST}$ most massive stars. All three clusters go through periods of significant mass segregation. The state of mass segregation is also highly variable throughout the cluster lifetime; mass segregation is not monotonic. M4 becomes mass segregated from $t_{\rm cl}\approx$~0.75--1.25~Myr for $N_{\rm MST}\leq7$ and mildly so from $t_{\rm cl}\approx$~1.75--2.4 for $N_{\rm MST}\leq10$. At the end of these time periods, the cluster transitions sharply to an inverse mass segregated state with $\Lambda<1$. From the bottom half of the M4 plot we can see that this is due to a new most-massive star being formed on the outskirts of the cluster, reducing $\Lambda$ for all values of $N_{\rm MST}$. M4 achieves a stable state of mass segregation for $N_{\rm MST}\leq5$ by $t_{\rm cl}\approx3.7\rm~Myr$. This is after the collapse of the cluster, the point of maximum compression where the sub-clusters have merged to form a short-lived dense core. The dense core begins to expand as the collapsed cluster relaxes. The onset of gas expulsion also contributes to cluster expansion. In lower star formation efficiency (SFE) clouds (SFE$\leq30\%$) gas expulsion is the dominant cause of expansion, while higher SFE clusters expand due to stellar dynamics \citep{Pfalzner2013A&A...559A..38P}. The M4 cluster expansion is driven by gas expulsion, whereas the M5 and M6 cluster expansion is due to dynamical relaxation. 

M5 has similar episodes, with mass segregation from $t_{\rm cl}\approx$~0.35--0.5 for $N_{\rm MST}\leq4$ and $t_{\rm cl}>1.9\rm~Myr$ for $N_{\rm MST}\leq20$ after collapse. Just before the onset of mass segregation at times $t_{\rm cl}=0.35\rm~Myr$ and $t_{\rm cl}=1.9\rm~Myr$ a new most massive star formed. Unlike in the M4 case, these massive stars formed near the center of mass of the cluster, thereby increasing $\Lambda$. 
This is dynamically induced primordial mass segregation. As the cluster assembles, dense gas is also pulled to the center of mass allowing for the formation of a massive star in the center of the cluster. This is more likely during collapse when segregation is most rapid, as is seen with the massive star that formed at $t_{\rm cl}=1.9\rm~Myr$ just after collapse. 
After $t_{\rm cl}=2\rm~Myr$, all stars $N_{\rm MST}\leq100$ show signs of mass segregation with the exception of $N_{\rm MST}=$3--7 where $\Lambda<1$. This region of inverse mass segregation is caused by an interaction at time $t_{\rm cl}=2.1\rm~Myr$ in which the fourth most massive star ($N_{\rm MST}=4$) is dynamically ejected from the cluster's core. Then, at $t_{\rm cl}=2.27\rm~Myr$ the third and fourth most massive stars swap places due to the third star losing mass from stellar winds. The ejected massive star stays in the outskirts of the cluster for the remainder of the simulation, affecting $\Lambda$ values for $N_{\rm MST}=$~3--7. 

The M6 cluster begins to show signs of mass segregation for $N_{\rm MST}\geq3$ as early as $t_{\rm cl}=0.2\rm~Myr$ after formation. An episode of mass segregation for $N_{\rm MST}\leq20$ occurs from $t_{\rm cl}=$~0.56--0.68 Myr, which is just after collapse. This ends when the second most massive star is dynamically ejected from the core, decreasing all values of $\Lambda$. The higher stellar mass of the M5 and M6 cluster results in a denser core and therefore more dynamical ejections of the most massive stars. The most massive stars are more susceptible to these ejections as dynamical friction drags them more quickly to the center of the core. This is one of the reasons why runaway stars---stars unbound to their birth cluster \citep[see, e.g.,][]{Blaauw1961BAN....15..265B,Poveda1967BOTT....4...86P,Hoogerwerf2000ApJ...544L.133H, Fujii2011Sci...334.1380F}---are preferably massive OB-type stars \citep{Oh2016A&A...590A.107O}.
\responsetwo{Although some massive stars form stochastically in the central cluster region, the dominant process driving mass segregation during core collapse is stellar dynamics. This is demonstrated in Appendix~\ref{app:old_star_ms} where we calculate $\Lambda$ for only old stars formed before $0.75~t_{\rm collapse}$. We find that for a fixed mass population, older stars become more mass segregated than all stars, including those formed during collapse in the most massive cluster, M6, where the IMF is best sampled. Older stars in the M4 and M5 clusters show comparable levels of mass segregation to the total population.}

While star formation continues and massive stars remain young, $\Lambda$ can be highly variable because new massive stars are forming throughout the cluster and are losing mass through stellar winds. To remove some of the dependence of $\Lambda$ on individual stars, we also calculated the average $\Lambda$ for all stars above a given stellar mass $M_\Lambda$, which is shown in Figure~\ref{fig:msr}. We only calculate $\Lambda$ if there are $N_{\rm mass}\geq5$ stars with mass $\geq M_\Lambda$. If there are $N_{\rm mass}\geq500$ stars above a mass, we set $N_{\rm mass}=500$ and average the MST length of 100 sets of randomly selected $N_{\rm mass}=500$ massive stars to increase computational efficiency.

The M4 cluster shows immediate mass segregation for stars above $20~\rm M_\odot$ which sharply transitions towards uniformity/inverse mass segregation just after $t_{\rm cl}=2.4\rm~Myr$. This is the same feature seen in Figure~\ref{fig:Nmst} when a new most massive star formed in the outskirts of the cluster. $\Lambda$ values in M4 begin to increase towards uniformity after collapse. M5 has some fluctuations in the segregation of the high mass stars, with a prominent peak at $t_{\rm cl}=1\rm~Myr$ for stars above $60\rm~M_\odot$. This peak occurs after collapse, when the dense core drove efficient migration of the massive stars to the center.
For the next megayear, $\Lambda$ for stars more massive than $50\rm~M_\odot$ shows consistent mass segregation. Higher mass stars above $70\rm~M_\odot$ evolve to be more uniform/slightly inverse mass segregated, though. This is a result of dynamical ejection, which acts on the most massive stars in a cluster as they reach the highest levels of mass segregation. The higher mass cutoffs also include many fewer stars in the calculation, meaning they are more sensitive to one star being ejected from the cluster core. 

Stars below $80\rm~M_\odot$ in the M6 cluster show signs of little to no mass segregation towards the end of the simulation $t_{\rm cl}\geq0.55\rm~Myr$ when the cluster starts to relax. Before this, $\Lambda$ for stars below $80\rm~M_\odot$ is highly variable due to the high star formation rate. This is also before the sub-clusters have merged, so global mass segregation can not take place. 

As in the M5 cluster, $\Lambda$ in the M6 cluster has peaks and subsequent dips. However, there are more peaks and dips because M6 has more sub-cluster mergers before they all coalesce during collapse. The peaks before collapse are caused by similar mass sub-cluster mergers. The final peak after collapse occurs at $t_{\rm cl}=0.56\rm~Myr$ for stars above $80\rm~M_\odot$, and afterwards $\Lambda$ levels off at $\Lambda=1.2$ indicating clear mass segregation. The most massive stars in M6, however, show less mass segregation after collapse due to their ejection from the core. The high concentration of the most massive stars in the dense core raises the probability of their ejection through a strong dynamical encounter. The low number of stars with $M \ge 90\rm~M_\odot$ results in $\Lambda$ dipping below one if even a single star in this mass bin is ejected from the core. 

\begin{figure}[t]
\centering
    \includegraphics[width=1.0\hsize]{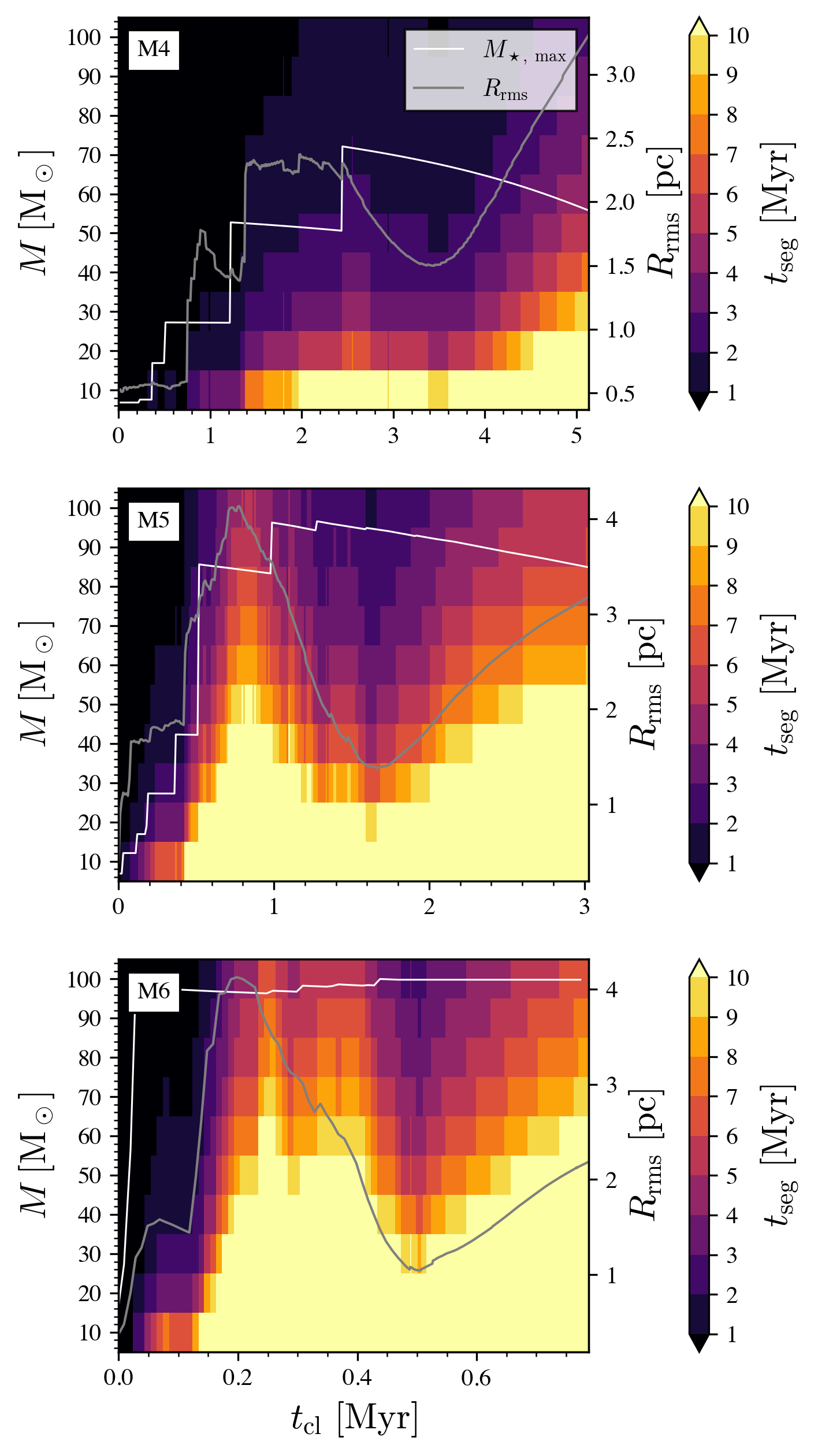}
    \caption{Dynamical segregation time for a star with mass $M$ over time based on the cluster's properties. The white line indicates the mass of the most massive star, which can decrease due to mass loss from stellar winds. The grey line shows the evolution of the cluster radius $R_{\rm rms}$. Note that the dip in the evolution of $R_{\rm rms}$ indicates the time of collapse. }
    \label{fig:t_seg}
\end{figure}

\section{Discussion}
\label{sec:discussion}
\subsection{Timescales}
An observed star cluster is determined to have primordial mass segregation if there is mass segregation before the expected segregation timescale of the cluster. A recent study that consistently models individual star formation from gas found that sub-clusters form primordially segregated \citep{Gusnejnov2022MNRAS.515..167G}. They do not indicate $\Lambda$ values for individual stars using $N_{\rm MST}$, but rather the global $\Lambda$ for all stars $\geq 5\rm~M_\odot$. Despite the primordial mass segregation in the sub-clusters, the global $\Lambda$ still varies as their cluster model evolves. This is in contrast to the results in \citet{Allison2009ApJ...700L..99A} who found that $\Lambda$ mostly increases after collapse, never decreasing back down to $\Lambda=1$. \citet{Mcmillan2007ApJ...655L..45M} also found in their pure N-body simulations that $\Lambda>1$ values in sub-clusters are retained through mergers. These conflicting results suggest that gas dynamics play a significant role in the long term relaxation and dynamical mass segregation of the cluster.

We calculate the dynamical segregation timescale for our clusters to see when mass segregation occurs with respect to it. The expected time for a star of mass $M$ to dynamically segregate in a uniform spherical cluster is given by \citep{Spitzer1969ApJ...158L.139S, Allison2009ApJ...700L..99A}
\begin{equation}
\label{eq:tseg}
    t_{\rm seg}\approx\frac{\langle m_\star \rangle}{M}\frac{N_\star}{8\log N_\star}\frac{R_{\rm rms}}{\sigma_{\bf\rm v}}\;,
\end{equation}
where $\langle m_\star \rangle$ and $N_\star$ are the average mass and number of stars in the cluster, $R_{\rm rms}=\sqrt{\langle ||\Delta \textbf{\rm x}||^2 \rangle}$ is the root mean square radius from the center of mass, and $\sigma_{\bf\rm v}$ is the stellar velocity dispersion in the center of mass frame. We note that this is an idealized approximation of the segregation time, as our clusters are far from spherical at early times.

We must address the effect of our mass agglomeration routine on the segregation timescale in our clusters. We agglomerated stars below $4\rm~M_\odot$. With an IMF sampling mass of $10^5\rm~M_\odot$, our agglomeration routine increases $\langle m_\star \rangle$ from 0.575 to 5.390~M$_\odot$ and decreases $N_\star$ from 174,432 to 18,243. Using Eq.~\ref{eq:tseg}, we find that the segregation time of a cluster with agglomeration is $20.6\%$ larger than the segregation time for a cluster without agglomeration. This is an approximation that assumes that agglomeration does not change the radius of the cluster or stellar velocity dispersion. This result means that clusters with the true Kroupa IMF will dynamically mass segregate even more efficiently than our model with the agglomerated IMF.
% N = 174432
% M_ave = 0.5751728002022599
% N agg = 18243
% M_ave agg = 5.389787859730862
% t_agg factor= 1.2055580509484596
\response{The vast majority of the analysis in this paper calculates $\Lambda$ for stars with masses $\geq10~\rm M_\odot$, while agglomerated stars have mass $<8~\rm M_\odot$. Because most stars $<10~\rm M_\odot$ are agglomerated, in a sense measuring mass segregation in our model means comparing the distribution of massive stars to low-mass agglomerate stars. The presence of mass segregation in all our models means that there are more agglomerated stars outside the central cluster region than inside. Therefore, even if we broke up the agglomerated stars into their constituents, our clusters would still have $\Lambda>1$ because the number of low mass stars would still be greater outside the central cluster region.}

Figure~\ref{fig:t_seg} shows the time evolution of $t_{\rm seg}$ as a function of $M$ for each of the modeled clusters. The mass of the most massive star is indicated by the white line and the grey line shows the evolution of $R_{\rm rms}$. In all three clusters, there is a distinct wave pattern of high and low $t_{\rm seg}$ values which correspond to the variations in $R_{\rm rms}$ during cluster assembly.
Over time, $N_\star$ and $\sigma_{\bf\rm v}$ only increase and $\langle m_\star \rangle$ is roughly constant by construction. The segregation time increases as more pockets of dense gas begin to form spatially separate sub-clusters of stars, and decreases when the sub-clusters are pulled together by gravity. The lowest $t_{\rm seg}$ value corresponds to the maximum contraction of the assembling cluster during collapse, after which the cluster expands and $t_{\rm seg}$ increases again. 

Comparing Figure~\ref{fig:t_seg} to Figure~\ref{fig:Nmst}, one can see a direct correlation between the episodes of high mass segregation to low segregation times, particularly the time period just after the dense core forms from collapse. This is the mass segregation mechanism described in \citet{Allison2009ApJ...700L..99A} in which the collapse of a star cluster forming from many sub-clusters creates a short-lived dense core that allows for early and efficient dynamical mass segregation in young clusters. They discovered this mechanism with models of $N=1,000$ star particles initially distributed in sub-clusters. Our model confirms that these results hold for sub-clusters forming self-consistently from the dense substructures in a collapsing cloud. One key difference in our results is that they found $\Lambda$ to steadily increase after collapse, whereas our clusters show more variability. We suspect this is due to the gas dynamics in our model slowing the relaxation of the star particles. 

Starting with the M4 cluster, the first episode of mass segregation is from $t_{\rm cl}\approx$~0.8--1.2~Myr, which occurs before $t_{\rm seg}$ begins to increase as more stars form. The final period of steady mass segregation begins at $t_{\rm cl}\approx3.7\rm~Myr$, which is right after the period of low $t_{\rm seg}$ due to the collapse of the cluster. This allows the most massive stars to efficiently reach a state of mass segregation. 

The M5 cluster also has an episode of mass segregation when the cluster is young $t_{\rm cl}\approx0.4\rm~Myr$ and just after the collapse when $t_{\rm seg}$ is lowest.
The collapse induced mass segregation occurs at $t=$~1.8--2~Myr, just $0.2\rm~Myr$ after the maximum contraction of the cluster. By the time mass segregation sets in, $t_{\rm seg}$ has risen considerably. At $t_{\rm cl}=1.95\rm~Myr$, there is pronounced mass segregation of $\Lambda\geq2$ for $N_{\rm MST}\leq20$ which corresponds to stellar masses between $M=$~50--100~M$_\odot$. The segregation timescale for stars in this mass range is between $t_{\rm seg}=$~2--6~Myr. For stars between $M=$~50--70~M$_\odot$, $t_{\rm seg}=$~3--6~Myr which is greater than the age of the cluster. If observing this cluster at that time, one would mistakenly conclude that the massive stars have segregated before the expected dynamical time and therefore must have been primordially segregated. This can occur if observing the cluster just after the initial collapse. 

The star formation in M6 is too rapid for mass segregation to set in before multiple sub-clusters form. However, the initial collapse does produce an episode of mass segregation from $t_{\rm cl}=$~0.56--0.68~Myr. The minimum $t_{\rm seg}$ values due to collapse occurs at $t_{\rm cl}=0.5\rm~Myr$. The two most massive stars are segregated before the collapse, and just $0.06\rm~Myr$ after the maximum contraction the 20 most massive stars become mass segregated as well. The 20 most massive stars ($N_{\rm MST}\leq20$) have $M\geq90\rm~M_\odot$. The segregation timescale for these stars during the time period of segregation ranges from $t_{\rm seg}=$~3--6~Myr. This is again much older than the age of the cluster.

\begin{figure}[t]
\centering
    \includegraphics[width=1.0\hsize]{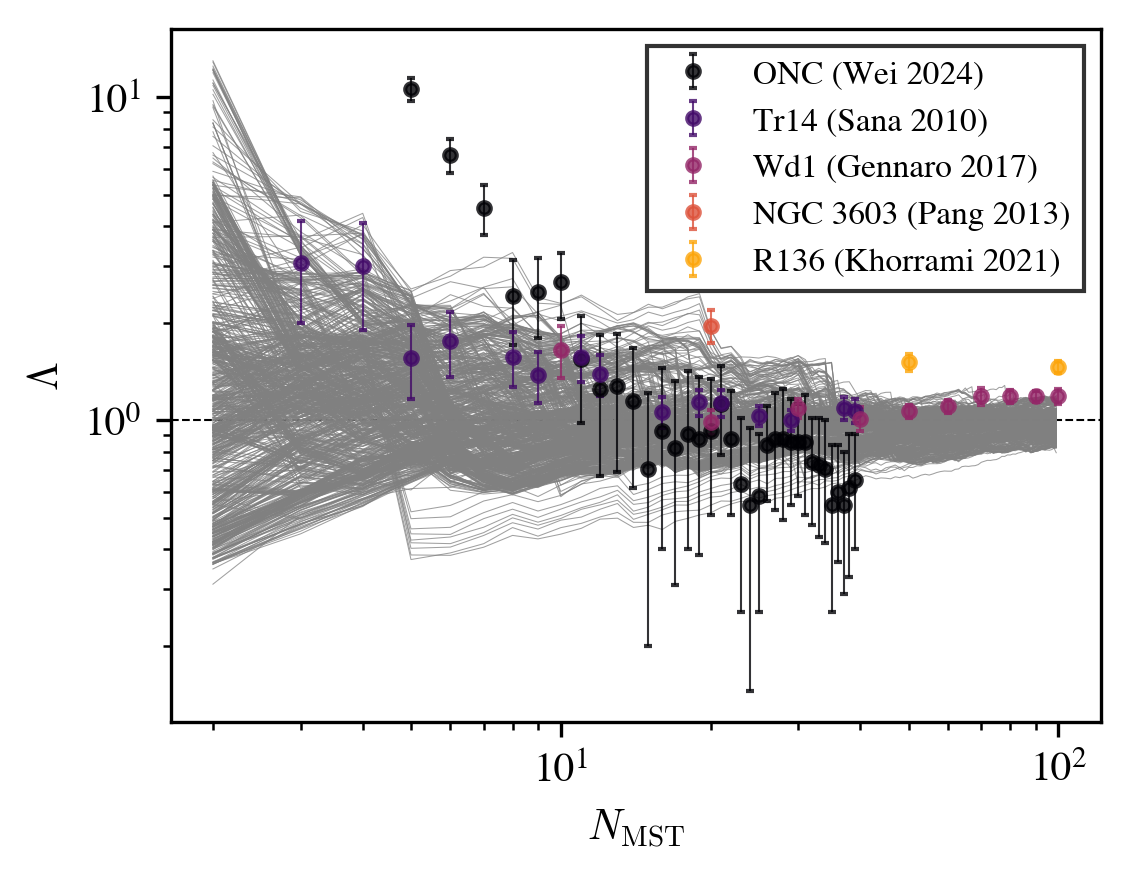}
    \caption{Mass segregation ratio $\Lambda$ versus number of most massive stars $N_{\rm MST}$ for young Galactic clusters. Each grey line represents one snapshot in time for each of the M4, M5, and M6 clusters.}
    \label{fig:obs}
\end{figure}

\subsection{Observational comparisons}
The mass segregation ratio $\Lambda$ for five young clusters in the Milky Way is shown in Figure~\ref{fig:obs}, with the values for our three models indicated in grey. This allows us to assess whether the state of mass segregation observed in these clusters can reasonably be attained with only dynamics. Due to the chaotic nature of star formation and dynamics, we are not expecting an exact match in $\Lambda$ for a given $N_{\rm MST}$ value. Rather, we are looking to see whether the $\Lambda$ values in our models reach those seen in observations for a given $N_{\rm MST}$. For reference, the cluster mass and age of our models and the Milky Way clusters are listed in Table~\ref{table:obs}.

The degree of mass segregation in the Trumpler 14 \citep[Tr14; ][]{Sana2010A&A...515A..26S}, Westerlund 1 \citep[Wd1; ][]{Gennaro2017MNRAS.472.1760G}, and NGC 3603 \citep{Pang2013ApJ...764...73P} clusters are well within the $\Lambda$ values of our simulated clusters for the same $N_{\rm MST}$. Each of these studies concluded that the mass segregation of these clusters can be explained by dynamics and does not have to be primordial because of their segregation timescale. Our results confirm this.

The young massive cluster Radcliffe 136 (R136, see, e.g., \citealt{Massey1998} or \citealt{Crowther2010}) embedded in the 30 Doradus star-forming region is mass segregated with $\Lambda\approx1.5$ at $N_{\rm MST}=50,~100$ \citep{Khorrami2021MNRAS.503..292K}. At times, our models are slightly segregated at these $N_{\rm MST}$, with the maximum values reaching $\Lambda\approx1.1-1.2$. \citet{Khorrami2021MNRAS.503..292K} notes that the detection completeness is very low for low-mass stars in the center of the cluster. A young massive cluster such as R136 is likely to have a dense stellar core with many low-mass stars. More low-mass stars in the center of the cluster would lower $\Lambda$, so it is probable that these values of $\Lambda$ are upper limits for R136. With this we conclude that the mass segregation in R136 can be reasonably attained through early dynamics.

All values of $\Lambda$ in the Orion Nebula Cluster \citep[ONC; ][]{Wei2024ApJ...962..174W} are reached by our model except for the massive Trapezium stars at $N_{\rm MST}\leq7$. Due to the high multiplicity fraction of massive stars \citep{Moe2017ApJS..230...15M}, it highly likely that the Trapezium stars are binaries rather than single stars. The binaries would undergo mass segregation as a system, increasing $\Lambda$ for two stars instead of one. We do not include primordial binary formation in our model, which can explain why our models reach the high $\Lambda$ values seen in the ONC at half the value of $N_{\rm MST}$.

The \citet{Allison2009ApJ...700L..99A} model, however, did reproduce the mass segregation seen throughout the ONC dynamically. Because of the stochastic nature of star formation and sub-cluster dynamics, our results will vary with different realizations of the initial turbulent seed of the clouds. It is likely that by performing more simulations we would produce a group of $N_{\rm MST}=7$ highly centralized stars resembling the Trapezium group. 

\citet{Wei2024ApJ...962..174W} find the ONC to be in a super-virial state and expanding. If the ONC formed through hierarchical assembly, the expanding state of the ONC implies that it is post-collapse. This is consistent with the scenario that the ONC became dynamically mass segregated during initial collapse, and is now expanding\footnote{It has also been suggested that the super-virial state of the ONC might be due to a slingshot effect, where oscillations in the gas filament that formed the ONC led to the ejection of the cluster \citep[see ][]{Stutz2016A&A...590A...2S, Stutz2018MNRAS.473.4890S, Matus2023MNRAS.522.4238M}. In this scenario, it is still possible that the initial collapse of the ONC occurred before expulsion by the filament, allowing for the Trapezium stars to efficiently migrate to the center.}. It is also possible that some of the Trapezium stars formed in the center of the ONC if their progenitor gas clumps became mass segregated during collapse.

\begin{table}
\caption{Star cluster properties.}   
\label{table:obs}   
\centering   
\begin{tabular}{l|cc}
\hline\hline                        
 & $M_{\rm cl}\rm~[M_\odot]$ & Age~[Myr] \\ 
\hline 
    M4 & $3.6\times10^3$ & $5.5 (3.1)$ \\
    M5 & $6.5\times10^4$ & $3.1 (1.9)$ \\
    M6 & $8.9\times10^5$ & $0.8 (0.4)$ \\
\hline
    ONC & $1.8\times10^3$ & $2.2$\\
    Tr14 & $4.3^{+3.3}_{-1.5}\times10^3$ & $0.3-0.5$ \\
    Wd1 & $5\times10^4$ & $4-5$ \\
    NGC3603 & $10^4$ & $1$ \\
    R136 & $1.5\times10^4$ & $1-2$ \\
\hline  
\end{tabular}
\tablefoot{Final age and stellar mass of our modeled clusters and the Galactic clusters shown in Figure~\ref{fig:obs}. We also list the average stellar age of our modeled clusters in parentheses. ONC mass \citet{Hillenbrand1998ApJ...492..540H}, ONC age \citet{Reggiani2011A&A...534A..83R}, NGC3603 mass \citet{Harayama2008ApJ...675.1319H}. All other observed values are taken from the papers reporting their mass segregation.}
% \tablefoot{R136 \citep{Khorrami2021MNRAS.503..292K}; ONC \citep{Wei2024ApJ...962..174W}; Wd1 \citep{Gennaro2017MNRAS.472.1760G}; NGC 3603 \citep{Pang2013ApJ...764...73P}; Tr14 \citep{Sana2010A&A...515A..26S}.}
% M4 age= 5.464008297348734 Myr , mass= 3623.888840372616 Msun
% M5 age= 3.141964142576559 Myr , mass= 64597.48117210678 Msun
% M6 age= 0.7906823310161901 Myr , mass= 891852.5064731997 Msun
\end{table}

Up to now, we have compared observations to our model, which has no primordial mass segregation by construction. We find that observations of early mass segregation can be reproduced with no primordial mass segregation. Therefore, to confirm the existence of primordial mass segregation, massive stars must be observed in the center of a cluster as they are forming. Recent ALMA observations of 11 dense protoclusters ($M_{\rm clump}\geq10^3\rm~M_\odot$) find significant mass segregation of their prestellar and protostellar cores \citep{Xu2024ApJS..270....9X}. This indicates that stars forming within sub-clusters could be primordially mass segregated. If this is the case, as they merge during the collapse of the cluster we expect the mass segregation will only increase. However, it is unclear whether primordially mass segregated sub-clusters remain that way until they merge. 
\response{Though we do not include pre-main sequence evolution in our models, we see transient periods of dynamical mass segregation in sub-clusters shortly after they form and before they merge during cluster collapse, which are similar to the primordial segregation seen in the ALMA protoclusters.}
The concentration of massive stars in the center of a sub-cluster subjects them to a higher chance of being dynamically ejected. With these new results, sub-grid models of star formation should perhaps place massive stars preferentially towards the center of sub-clusters. Then one could determine whether primordial mass segregation persists through the dynamical evolution of the cluster. 
%mm [moved the end of this para up to the beginning of the section, as it seemed to fit better in the Timescales subsection as introductory material.]

\section{Conclusions}
\label{section:conclusions}

We performed simulations of star-by-star cluster formation from turbulent self-gravitating gas clouds, taking into account stellar feedback in the form of radiation, stellar winds, and SNe. We find that dynamical mass segregation can occur early on during the hierarchical formation process, when sub-clusters collapse and form a dense core with a much smaller crossing time (as proposed by \citealt{Allison2009ApJ...700L..99A} for purely stellar dynamical systems). 

Due to the hierarchical formation of star clusters, the timescale over which massive stars dynamically segregate $t_{\rm seg}$ varies greatly as the size of the cluster changes. 
There are two points in a cluster's lifetime when $t_{\rm seg}$ is significantly lower. First, when the cluster is just forming and only contains a single site of star formation, and second during the initial collapse when  the sub-clusters merge into a central cluster forming a short-lived dense core. These two time periods allow massive stars to segregate efficiently, with $t_{\rm seg}\leq2~\rm Myr$ for the most massive stars. During initial collapse, the most massive stars in each of our models ($\geq20\rm~M_\odot,~\geq50\rm~M_\odot,~\geq90\rm~M_\odot$ for M4, M5, M6) underwent efficient mass segregation reaching $\Lambda>1$ within $t_{\rm seg}<0.5\rm~Myr$. The variation of $t_{\rm seg}$ over the lifetime of a cluster means that the integrated $t_{\rm seg}$ may be shorter than the value given by the cluster's current state. Therefore, a young cluster being mass segregated earlier than expected does not necessarily indicate primordial mass segregation. 

Mass segregation can occur while star formation is ongoing. This means massive stars and massive star-forming clumps alike will migrate to the center, thereby increasing the likelihood of a massive star being born in the center of the cluster. This can lead to primordial mass segregation if a massive star is formed during the initial collapse of a young cluster. 

The mass segregation ratio $\Lambda$ in young clusters is highly variable due to the ongoing star formation, stellar mass loss, and energetic kinematics of sub-cluster assembly. There are time periods of significant mass segregation in all three clusters that last for 0.1--0.3~Myr. There are also time periods without apparent mass segregation, which usually occur after the dynamical ejection of a massive star from the core. As more massive stars fall into the dense core, they are more likely to be ejected by a strong dynamical encounter. Mass segregation and dynamical ejection are two competing physical processes, and massive stars are most susceptible to both. The interplay and balance of these two processes needs further investigation. Given these variations in $\Lambda$ before the cluster is fully relaxed, it is unclear how long primordial mass segregation would last.

In summary, our results conclusively show that young clusters can become mass segregated through early dynamics. Efficient dynamical mass segregation is achieved when the initial collapse of the star cluster forms a dense core with a much smaller crossing time. We have demonstrated that the degree of mass segregation seen in young Galactic clusters is also seen in our models and can therefore be achieved through dynamics alone. During collapse, massive stars are more likely to form primordially segregated as massive star-forming clumps will also promptly approach the center of mass. Our findings also indicate that $\Lambda$ is highly variable at early times. New observations point to primordial mass segregation, at least on the sub-cluster scale, so further work is needed to confirm whether primordial mass segregation would survive cluster assembly or dissipate in a model that includes gas dynamics.

\begin{acknowledgements}

\response{We thank the referee for their useful comments and insights, one of which led to the addition of Appendix~\ref{app:old_star_ms}.} B.P. was partly supported by a fellowship from the International Max Planck Research School for Astronomy and Cosmic Physics at the University of Heidelberg (IMPRS-HD). M.-M.M.L., B.P., and E.A. were partly supported by NSF grants AST18-15461 and AST23-07950. E.A. and M.-M.M.L. also acknowledge partial support from NASA grant 80NSSC24K0935. C.C.-C. is supported by a Canada Graduate Scholarship - Doctoral (CGS D) from the Natural Sciences and Engineering Research Council of Canada (NSERC). This work used Stampede 2 at TACC through allocation PHY220160 from the Advanced Cyberinfrastructure Coordination Ecosystem: Services \& Support (ACCESS) program, which is supported by National Science Foundation grants 21-38259, 21-38286, 21-38307, 21-37603, and 21-38296. The code development that facilitated this study was done on Snellius through the Dutch National Supercomputing Center SURF grants 15220 and 2023/ENW/01498863. S.M.A. is supported by an NSF Astronomy and Astrophysics Postdoctoral Fellowship, which is funded by the National Science Foundation under Award No. AST24-01740; S.M.A. also acknowledges support from NSF Award No. AST20-09679. R.S.K.\ and S.C.O.G.\ acknowledge financial support from the European Research Council via the ERC Synergy Grant ``ECOGAL'' (project ID 855130),  from the Heidelberg Cluster of Excellence (EXC 2181 - 390900948) ``STRUCTURES'', funded by the German Excellence Strategy, and from the German Ministry for Economic Affairs and Climate Action in project ``MAINN'' (funding ID 50OO2206). The team in Heidelberg also thanks {\em The L\"{a}nd} and the German Science Foundation (DFG) for computing resources provided in bwHPC supported by grant INST 35/1134-1 FUGG and for data storage at SDS@hd supported by grant INST 35/1314-1 FUGG. M.-M.M.L. acknowledges Interstellar Institute's program ``II6'' and the Paris-Saclay University's Institut Pascal for hosting discussions that nourished the development of the ideas behind this work.

\end{acknowledgements}

\bibliographystyle{aa}
\bibliography{references}{}
\onecolumn
\begin{appendix}

\section{Isolating dynamical mass segregation}
\label{app:old_star_ms}

\begin{figure*}[h]
\centering
    \includegraphics[width=0.95\textwidth]{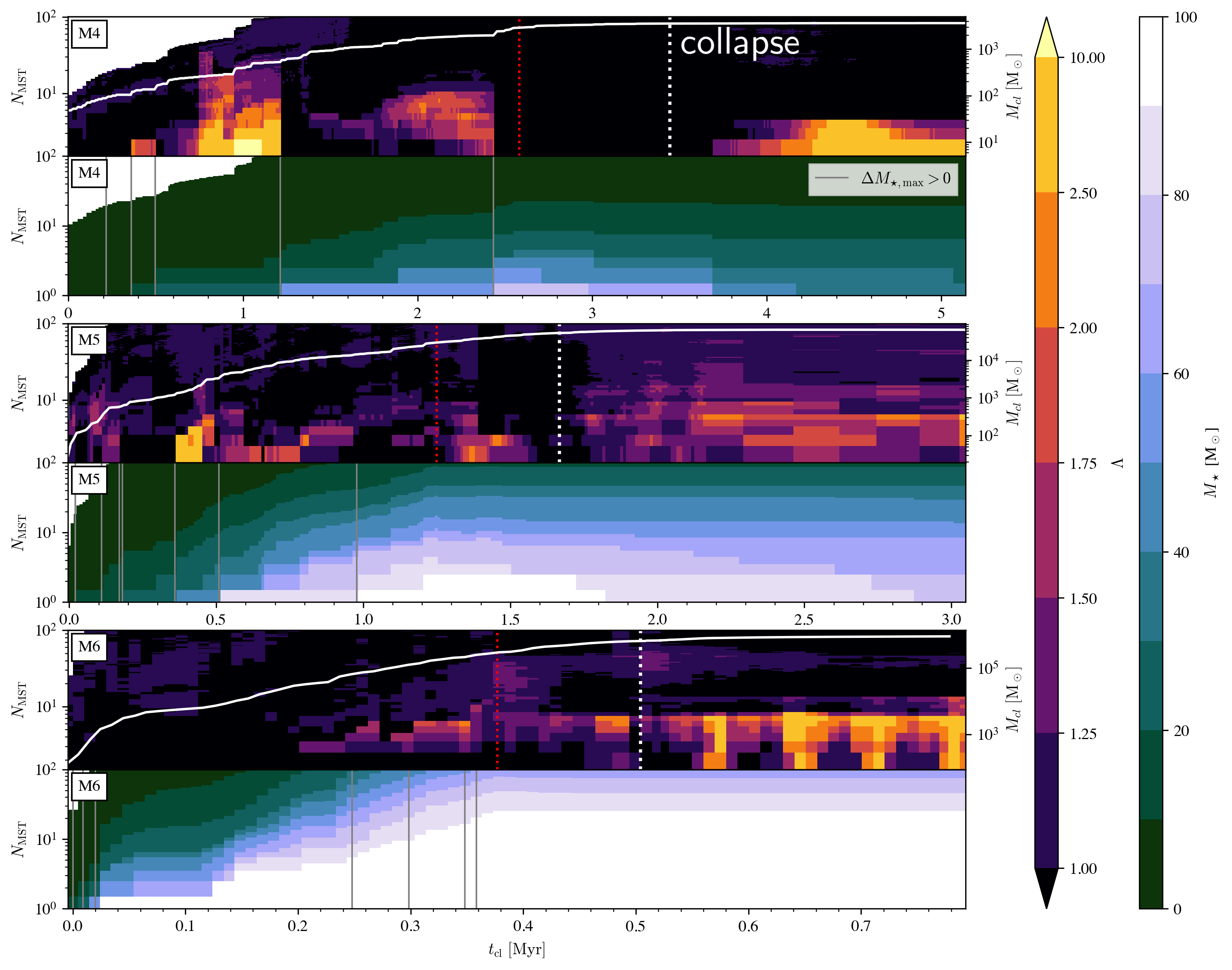}
    \captionof{figure}{\response{Mass segregation ratios $\Lambda$ and masses of $N_{\rm MST}$ most massive stars for an older subset of the stellar populations. Same as Figure~\ref{fig:Nmst}, but only considering stars formed before $0.75~t_{\rm collapse}$, indicated by the {\em red dotted line}. The three rows 
    %with two grouped plots
    correspond to the M4, M5, and M6 models as labeled. In each row: {\em (top)} mass segregation ratio $\Lambda$ over time for the $N_{\rm MST}$ most massive stars in each cluster. The vertical dotted white lines indicate the time of collapse, where $R_{\rm rms}$ reaches a minimum. The solid white lines correspond to the right vertical axis showing the stellar mass of the cluster. {\em (bottom)} mass of the $N^{\rm th}_{\rm MST}$ most massive star in each cluster. The grey vertical lines indicate the formation of a new most-massive star ($N_{\rm MST}=1$). As these are often not in the core, the time of formation, particularly in M4, corresponds to a drop in the apparent mass segregation. Note that each cluster was run to $\approx1.5t_\mathrm{ff}$, leading to different absolute timescales.}
    \label{fig:Nmst_old}}
\end{figure*}

Star formation continues during core collapse, so it must be determined whether ongoing star formation or stellar dynamics drives mass segregation during collapse. We do this by repeating the mass segregation analysis for only stars formed before $0.75~t_{\rm collapse}$.\ 
Comparing Figure~\ref{fig:Nmst_old} to Figure~\ref{fig:Nmst}, respectively, it is clear that dynamical mass segregation is occurring even without the aid of ongoing star formation placing massive stars in the center of the cluster. 

\responsetwo{
The mass segregation of old stars in M4 is almost identical to all stars. This is because the most massive star to form in M4 is created before the $0.75~t_{\rm collapse}$ cutoff. 
The old population in M5 becomes well segregated after collapse across all $N_{\rm MST}=100$ stars considered. A few stars more massive than any in the old population form after $0.75~t_{\rm collapse}$, so the high $\Lambda$ at low $N_{\rm MST}$ reached in the complete population is not seen when considering the old population. M4 and M5 both indicate that these clusters segregate dynamically without significant aid from stochastic massive star formation in the central cluster during collapse.

The older stars in M6 become much more segregated than the stars that are newly formed. M6 provides a better constraint for this test than M4 and M5, as the mass population of stars formed before and after $0.75~t_{\rm collapse}$ is comparable. This allows us to directly compare the degree of mass segregation for stars formed before and during collapse. The result that older stars in M6 become more mass segregated than young stars confirms that dynamics is the dominant mechanism segregating our young clusters during core collapse.}

\end{appendix}

\end{document}